# The Effect of Dipole Boron Centers on the Electroluminescence of Nanoscale Silicon $p^+$-n Junctions

Nikolay Bagraev[a], Leonid Klyachkin[a], Roman Kuzmin[a], Anna Malyarenko[a] and Vladimir Mashkov[b]

[a]*Ioffe Physical-Technical Institute, Polytekhnicheskaya 26, 194021, St.Petersburg, Russia*
[b]*St. Petersburg State Polytechnical University, Polytekhnicheskaya 29, 195251, St.Petersburg, Russia*

**Abstract.** Nanoscale silicon $p^+$-n junctions with very high concentration of boron, 5 $10^{21}$ cm$^{-3}$, are found to demonstrate interesting optical properties. Emission band dominated in near-infrared electroluminescence (EL) spectra possesses high degree of the linear polarization along preferred crystallographic axis which can be controlled by the lateral voltage applied in the plane of the $p^+$-n junction. Such behavior together with the temperature dependence of the EL intensity is attributed to the presence of the self-compensating dipole boron centers, $B^+$-$B^-$, with negative correlation energy which are identified using the ESR technique in the nanoscale silicon $p^+$-n junctions heavily doped with boron. The model of the recombination process though the negative-U dipole boron centers controlling the optical properties of the nanoscale silicon $p^+$-n junctions is proposed.

## INTRODUCTION

The studies of metastable centers with negative correlation energy are actual problem of the modern condensed matter physics. It was previously found that the shallow boron acceptors under high concentration demonstrate the negative-U properties and can reconstruct into the dipole centers [1,2]. In the present work we study the heavily boron doped nanoscale silicon $p^+$-n junctions which have been shown to represent the p-type silicon quantum well (Si-QW) confined by wide-band-gap heterobarriers on the Si (100) wafer [3]. The high concentration of boron, 5 × $10^{21}$ cm$^{-3}$, in the heterobarriers is essentially higher than the critical concentration of the Mott transition [4] which seems have to cause a metallic-like conductivity. Nevertheless, the angular dependences of the cyclotron resonance spectra demonstrated that the p-type Si-QW contains the high mobility gas of 2D holes which is characterized by long momentum relaxation time of the heavy and light holes [2, 3]. Thus, the momentum relaxation time of holes in the Si-QW appeared to be longer than in the best metal-oxide-semiconductor structures contrary to what might be expected from strong scattering by the heavily doped wide-band-gap heterobarriers. To eliminate this contradiction the electron spin resonance (ESR) technique was applied for the studies of the boron centers [2, 3]. The behaviour of the ESR spectra showed that the shallow boron acceptors are preferably form the trigonal dipole centers, $B^+$ - $B^-$, which are caused by the elastic reconstruction along the <111> crystallographic axes as a result of the negative-U reaction: $2B^o \rightarrow B^+ + B^-$.

In this work we study the effect of the trigonal dipole boron centers on the near-infrared electroluminescence of the nanoscale silicon $p^+$-n junctions. The electroluminescence is found to demonstrate the high degree of linear polarization along preferred crystallographic axes and can be controlled by the lateral voltage applied in the plane of the $p^+$-n junction.

## METHODS

The nanoscale silicon $p^+$-n junctions were prepared within frameworks of the planar diffusion silicon technology that relies on the short-time diffusion of boron from the gas phase after preliminary oxidation of the *n*-type Si (100) surface and subsequent photolithography processes.

Well-known generation of excess fluxes of the silicon self-interstitials and vacancies occurs during the formation of an oxide overlayer on the Si (100) wafer. These fluxes have the preferred crystallographic directions along <111> and <100> axes, respectively [5]. They allow the formation near the Si-SiO$_2$ interface the nanostructured layer of 8 nm wide produced by a set of self-interstitial microdefects confining longitudinal ultra-narrow Si-QW. After photolithography and etching the nonequilibrium short-time boron diffusion from the gas phase was used to passivate the dangling bonds and to transform the n-type Si (100) wafer with the nanostructured layer into the nanoscale silicon $p^+$-n junctions. The high concentration of boron introduced during this process was controlled by the SIMS method and equal to the value of $5 \times 10^{21}$ cm$^{-3}$. The confining potential inside the nanostructured layer heavily doped with boron was revealed by the cyclotron resonance quenching and the angular dependence for which characteristic $180^o$ symmetry was observed.

To perform electroluminescence (EL) measurements the gold contacts were prepared on the front surface of the $p^+$-n junction using standard photolithography technique, whereas the back side of the substrate was covered by aluminum. The device was designed within frameworks of the Hall geometry with the doping area of size $4.7 \times 0.1$ mm which is provided by eight contacts (Fig.1).

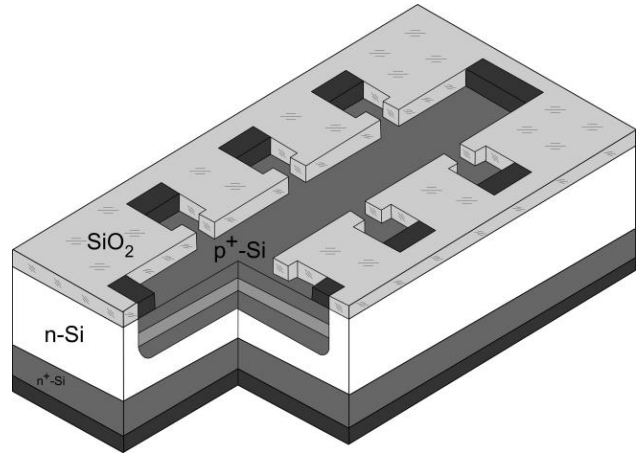

**FIGURE 1.** A schematic diagram of the device that demonstrates a perspective view of the nanoscale silicon $p^+$-n junctions which are the p-type nanostructured layer of 8 nm wide heavily doped with boron on the n-type Si (100) wafer. The nanostructured layer is produced by a set of self-interstitial microdefects confining the longitudinal ultra-narrow Si-QW.

## RESULTS AND DISCUSSIONS

The forward bias applied to the studied nanoscale silicon $p^+$-n junctions appears to result in the intensive near-infrared electroluminescence which is formed near the interface between the nanostructured layer and the n-type Si (100) wafer. The high concentration of boron was shown to play an important role in this electroluminescence spectra formation [6]. The emission band dominating in the spectra at $T > 20K$ seems to be either so-called HL or LL luminescence [7] depending on the initial state of recombination process in the n-type Si wafer. Where the final state is the boron acceptor band totally merging into the valence band [8].

The negative-U pairing and the reconstruction of shallow acceptors into the trigonal dipole centers are accompanied by a formation of the correlation gap in the boron acceptor band [3, 9]. The correlation gap distorts the density of states (DOS) which became different from its ordinary parabolic law. The distortion of the DOS involved into the recombination process appeared through the spectral line-shape changes of the near-infrared electroluminescence from nanoscale silicon $p^+$-n junctions. The changes were revealed at temperatures under $T \approx$ 146 K [10]. This temperature coincides with the critical temperature of the superconductor transition previously found in the nanoscale silicon $p^+$-n junctions [11]. The superconductivity was accounted for by transferring of the small hole bipolarons through the negative-U dipole boron centers. So the temperature increasing up to the critical value seems to suppress the charge correlations. It appeared in the EL measurements through the change of the DOS energy dependence which becomes closer to the square root law expected for the highest impurity concentrations [8].

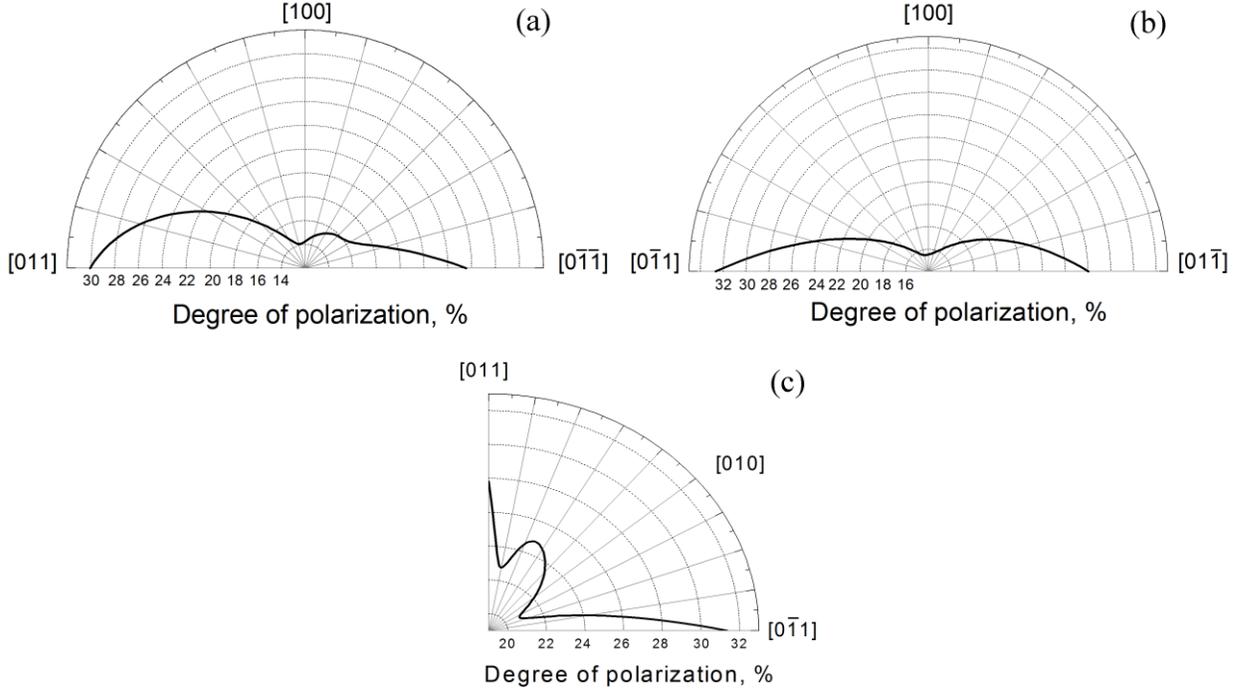

**FIGURE 2.** (a, b, c) Angular dependences of the degree of linear polarization of the electroluminescence from the nanoscale silicon p$^+$-n junctions heavily doped with boron which have been obtained during sample rotation in three perpendicular planes. The measurements were performed at the temperature of 77 K.

The involvement of the dipole boron centers in the recombination process leads to the fact that the detected EL exhibits at low temperatures a high degree of linear polarization which reaches more than 30%. The high degree of linear polarization is probably caused by ordering of the boron dipole centers along <111> axes. The angular dependences of the degree of linear polarization (Figs.2a, b, c) measured at various sample orientations with respect to the observation axis confirm this assumption. The observed angular dependences revealed the same trigonal symmetry which was found by the studies of the angular dependences of the ESR spectra. We can see that the degree of linear polarization is maximal along the <110> axis and equivalent ones which correspond to the boron dipoles alignment direction projected onto the (100) plane. It should be noted that the degree of linear polarization in the intermediate angles between directions equivalent to <110> decreases approximately by a factor of $\sqrt{2}$ as it should be in the case of the projection of an electric field vector oscillating along the <110> axis onto the <010> axis (Fig. 2c). The asymmetry of the angular distribution of the degree of linear polarization can be explained for example by the imperfection of the boron dipole centers sublattice. It can be caused by incomplete reconstruction of shallow boron acceptors or by local stresses appearing during the samples preparation.

The high degree of linear polarization and the intensity of the electroluminescence can be suppressed by the lateral voltage applied in the plane of the silicon p$^+$-n junctions (Figs. 3a, b). The reason of such behavior seems to be a distortion in the system of the trigonal dipole boron centers.

The recombination process in nanoscale silicon p$^+$-n junctions under the forward bias appearing in the interface between the nanostructured layer and the n-type Si (100) wafer can be accounted for by the set of the following reactions

$$\begin{aligned} \mathrm{B}^+ + e &\rightarrow \mathrm{B}^0 + h\nu \\ \mathrm{B}^- + h &\rightarrow \mathrm{B}^0 \\ 2\mathrm{B}^0 &\rightarrow \mathrm{B}^+ + \mathrm{B}^- \end{aligned} \quad (1)$$

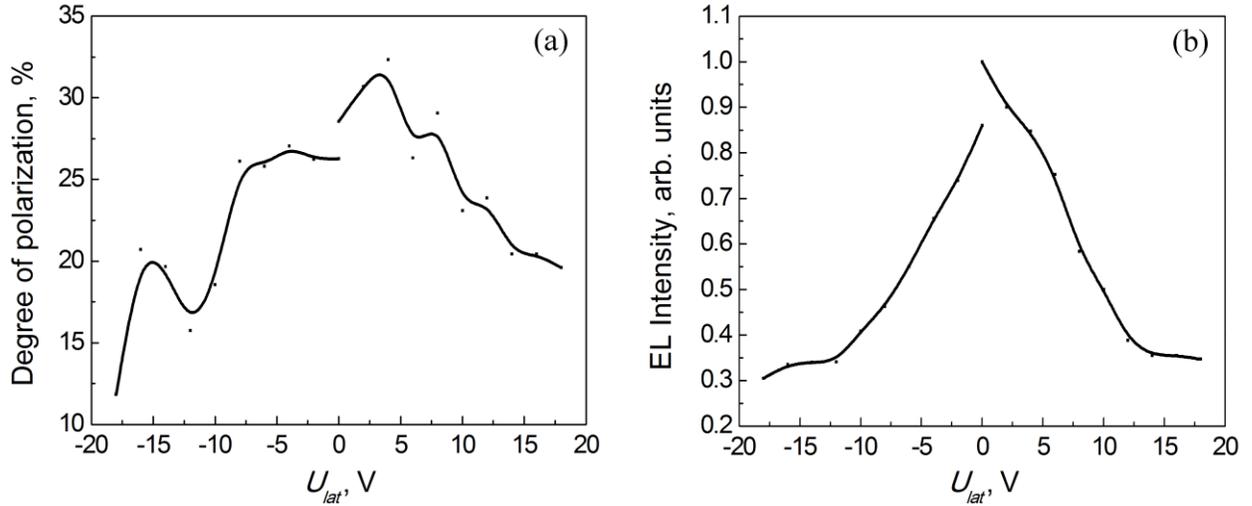

**FIGURE 3.** The degree of linear polarization (a) and intensity (b) of the electroluminescence from the nanoscale silicon $p^+$-n junctions heavily doped with boron which are quenched by the lateral voltage applied in the plane of the $p^+$-n junctions. The measurements were performed at the temperature of 77 K.

The electron transition from the states in the n-type Si wafer to the $B^+$ states in the $p^+$-nanostructured silicon layer results in the observed near-infrared electroluminescence. The Fig. 4a demonstrates this transition. The positions of the energy levels in the one-electron band diagram associated with $B^+$ and $B^-$ states (Fig. 4a) are determined by the energies of the ionization of the phone, $I_1$ - (0/+), and correlated, $\Delta I$ - (-/0), electrons that belong to the negative-U boron center (Figs. 4b, c). At the same time holes are captured by the $B^-$ states, and at least neutral boron acceptors reconstruct into the dipole boron centers. A great cross-section of this process caused by the negative-U properties of the center results in permanent depletion of the $B^0$-state thereby providing the high intensity of the electroluminescence. The presence of the preferred axis of the boron dipoles alignment appears to result in the linear polarization of the electroluminescence from the nanoscale silicon $p^+$-n junctions.

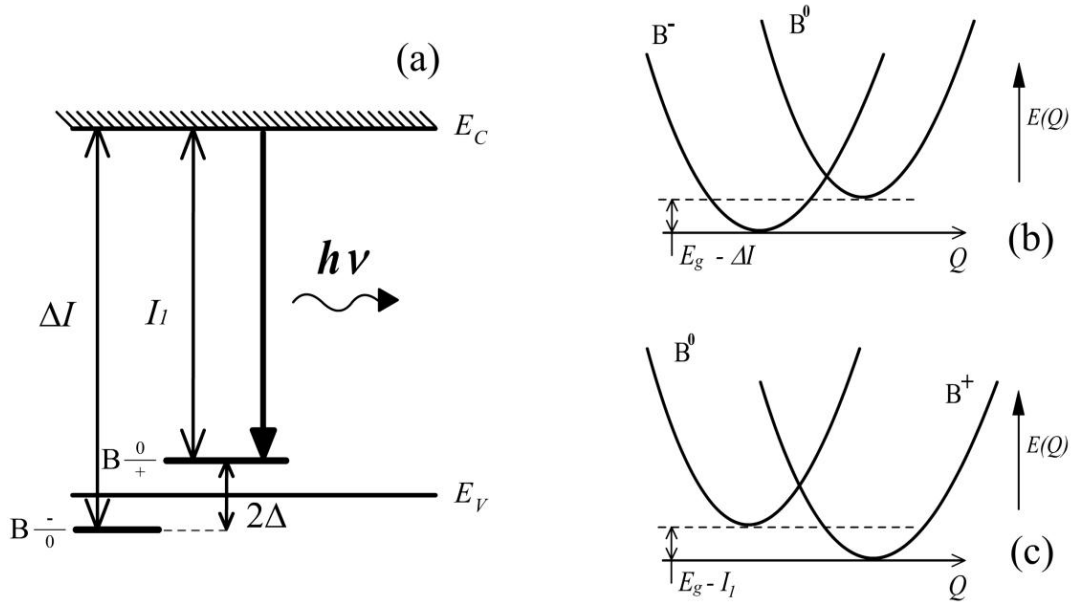

**FIGURE 4.** (a) The one-electron band diagram of the emission process due to recombination through the negative-U dipole boron centers. $I_1$ and $\Delta I$ are the energies of the ionization of the phone, (0/+), and correlated, (-/0), electrons that belong to the negative-U dipole boron center. The correlation gap, $2\Delta$, that is able to be transformed in the superconducting gap is equal to 0.044 eV. (b, c) Two-electron adiabatic potentials for the boron acceptor center in p-type silicon.

An analysis of the experimental data allows a conclusion that the effect of lateral voltage on the EL intensity and on the degree of its linear polarization seems to be caused by the current of holes flowing through the $p^+$ silicon nanostructured layer. The current of holes is due to the difference between the quasi-Fermi levels created by the lateral voltage applied in the plane of the $p^+$-n junction. The extra holes are captured by the $B^-$ states which disbalance the reactions (1) producing excess amounts of neutral boron acceptors. It should to result in decreasing of the intensity of the electroluminescence from the nanoscale silicon $p^+$-n junctions and in the suppression of the high degree of its linear polarization.

It is interesting to note a hysteresis in the dependences of the degree of the linear polarization and intensity of the electroluminescence on the lateral voltage (Figs. 3a, b). Such behavior appears to be started from incomplete reconstruction of some dipole boron centers after perturbation relieving. Nevertheless, this question seems to need further studying.

## CONCLUSIONS

The near-infrared electroluminescence from the nanoscale silicon $p^+$-n junctions is originated near the interface between the nanostructured layer and the n-type Si (100) wafer. It arises as a result of the recombination via boron acceptors which reconstruct into the trigonal dipole centers. This appears through the spectral line shape changes, high degree of the linear polarization of the electroluminescence and possibility to control it by varying the lateral voltage applied in the plane of the $p^+$-n junctions. The proposed model of the recombination process though the negative-U dipole boron centers seems to account for the observed behavior of the electroluminescence from the nanoscale silicon $p^+$-n junctions.

## ACKNOWLEDGMENTS

The work was supported by the programme of fundamental studies of the Presidium of the Russian Academy of Sciences "Quantum Physics of Condensed Matter" (grant 9.12); programme of the Swiss National Science Foundation (grant IZ73Z0_127945/1); the Federal Targeted Programme on Research and Development in Priority Areas for the Russian Science and Technology Complex in 2007–2012 (contract no. 02.514.11.4074), the SEVENTH FRAMEWORK PROGRAMME Marie Curie Actions PIRSES-GA-2009-246784 project SPINMET.